\documentclass[showpacs,preprintnumbers,amsmath,amssymb, preprint, prb]{revtex4}

\usepackage{graphicx}
\usepackage{dcolumn}
\usepackage{bm}
\usepackage{epsfig}
\begin{document}

\title{Quantum mechanical sum rules for two model systems}

\author{M. Belloni} \email{mabelloni@davidson.edu}
\affiliation{%
Physics Department \\
Davidson College \\
Davidson, NC 28035 USA \\
}

\author{R. W. Robinett} \email{rick@phys.psu.edu}
\affiliation{%
Department of Physics\\
The Pennsylvania State University\\
University Park, PA 16802 USA \\
}

\date{\today}

\begin{abstract}
Sum rules have played an important role in the development of many branches
of physics since the earliest days of quantum mechanics.
We present examples of one-dimensional quantum mechanical sum rules
and apply them in two familiar systems, the infinite well and the single
$\delta$-function potential. These cases
illustrate the different ways in which such sum rules can be realized,
and the varying mathematical techniques by which they can be confirmed.
Using the same methods, we also evaluate the second-order energy shifts
arising from the introduction of a constant external field, namely the
Stark effect.
\end{abstract}
\pacs{03.65.-w, 03.65.Db, 11.55.Hx}

\maketitle

\section{Introduction}

Quantum mechanical identities which relate time-dependent expectation values,
influential in the early days of quantum theory, continue
to act as useful pedagogical tools in the modern curriculum. For example,
the results often known as Ehrenfest's theorem(s),\cite{ehrenfest}
\begin{equation}
\langle \hat{p} \rangle_t = m \frac{d \langle x \rangle_t}{dt}
\quad
\mbox{and}
\quad
m \frac{d^2 \langle x\rangle_t}{dt^2}
= - \left\langle \frac{dV(x)}{dx}\right\rangle_t
\end{equation}
can be used to show that time-dependent quantum expectation values are related
to their corresponding classical equations of motion.\cite{styer}
Identities restricted to time-independent expectation values evaluated
using energy eigenstates, $|n\rangle$, such as the quantum virial theorem
\begin{equation}
\langle n |\hat{T}| n \rangle
=
\left\langle n\left| \frac{\hat{p}^2}{2m} \right| n \right\rangle
= \frac{1}{2}\left\langle n \left| x\, \frac{dV(x)}{dx}\right| n \right
\rangle
\end{equation}
and related hypervirial theorems,\cite{hypervirial} are
historically and pedagogically valuable as they too have clear
classical analogs and can often be evaluated without resorting to direct integration.

Similar relationships involving off-diagonal matrix elements,
especially quantum mechanical sum rules, were also used to dramatic
effect in the early days of quantum theory. For example, the
Thomas-Reiche-Kuhn (TRK)  energy-weighted sum rule,\cite{trk_sum_rule}
again for energy eigenstates,
\begin{equation}
\sum_{k} (E_k - E_n) |\langle n |x| k \rangle|^2
= \frac{\hbar^2}{2m}
\qquad
\mbox{(TRK sum rule)}
\label{trk_sum_rule}
\end{equation}
was used to describe the physics of electric-dipole interactions with
atoms.  It was originally obtained by requiring that the Kramers-Heisenberg dispersion
relation reduce to the Thomas scattering formula at high energies.
Written in the form
\begin{equation}
\sum_{k} \frac{2m(E_k - E_n)}{\hbar^2} |\langle n |x| k \rangle|^2
= \sum_{k} f_{n,k} = 1
\end{equation}
this was an important experimental check of the oscillator strengths
($f_{n,k}$) and an early confirmation of quantum results.
Kramer was able to derive this relation
in the context of matrix mechanics, reproducing the matrix version of
the famous commutation relation $[\,x,\,p\,] = i\hbar$.\cite{heisenberg_history}
Other early uses of sum rules included Bethe's study of energy loss
mechanisms for charged particles in matter,\cite{bethe_sum_rule} which
made use of the relation
\begin{equation}
\sum_{k}(E_k-E_n)
|\langle n |e^{iqx}|k\rangle|^2 = \frac{\hbar^2 q^2}{2m}
\;\mbox{(Bethe sum rule)}
\, ,
\label{bethe_sum_rule}
\end{equation}
eventually leading to the Bethe-Bloch formula.

Since then, sum rules have been used in many areas of physics,
including in
atomic,\cite{guiner}
molecular,\cite{shimamura}
solid state,\cite{van_hove,tinkham,tinkham_2,high_tc}
nuclear,\cite{isospin,bohigas,orlandini,cohen}
and especially in
particle physics.\cite{current_algebra,bjorken,narison,reinders,heavy_quark}
One well-known paper applying sum rule methods to
QCD \cite{shifman} is the 10$^{th}$ most highly cited paper in the particle
physics literature and over 2,000 papers on QCD sum rules
have been published, with 60 appearing in 2007 alone.\cite{citations}

The power of such sum rule identities is that they encode a
large amount of information about both the energy spectrum and energy
eigenfunctions of the system in a compact form, often in a way which is
amenable to experimental confirmation. This in turn can probe assumptions
about the fundamental interactions assumed or the calculation methods used
to approximate physical systems. For example, QCD sum rules have been
used to extract values of both the light and heavy quark masses,
which are not otherwise directly measurable quantities.

Despite their historical and contemporary importance, sum rules are not often treated
in the context of standard quantum mechanics courses. The TRK sum rule
is sometimes included in undergraduate quantum mechanics books,\cite{qm_books} but often only as a problem,
and typically only using the harmonic oscillator.
This lack of coverage may well be due to the paucity of tractable examples
in familiar model systems to which students typically have exposure, or
the level of mathematical analysis required to verify even the simplest
cases.

The purpose of this paper is to provide a suite of one-dimensional
sum rules and to demonstrate the mathematical techniques required
for their confirmation in two model quantum mechanical
systems, the  infinite well and the single
(attractive) $\delta$-function potential. {In each case, the sum rules
saturate differently and rely on different mathematical methods (summation
techniques and contour integration methods) illustrating the diverse ways
in which such sum rules are realized. The level of
mathematical detail required, however, is low enough to be
easily accessible to advanced undergraduate students.

Explicitly confirming that these identities are indeed satisfied is
not an empty exercise since it is possible to obtain surprising results,
even from  relatively simple systems such as the rigid rotator.\cite{trk_rotator}
In addition, energy-weighted sum
rule calculations are actually not exotic, since perturbation theory
is discussed in standard textbooks in quantum mechanics. The expression
for the second-order shift in the energy due to a
perturbation $V'(x)$ is given by
\begin{equation}
E_n^{(2)} = \sum_{k \neq n}
\frac{|\langle n |V'(x)|k\rangle|^2}{(E_{n}^{(0)} - E_{k}^{(0)})}
\label{second_order_shift}
\end{equation}
which is a form of energy-weighted sum rule.
Using this connection,
we will find that we can make use of the exactly same techniques derived
for confirming sum rules to evaluate the shift due to the addition of a
constant external field, $V'(x) = Fx$, namely the Stark shift, in each
of the model systems we consider.

Introducing the concept of sum rules,
using the tractable examples
considered here, can certainly help students appreciate their use in later
research applications. It can also help put the mathematical methods
used in the same context as more familiar second-order perturbation theory
calculations, one of the most important applications of undergraduate
quantum mechanics, showing how the related required sums over intermediate
states can be sometimes done in closed-form, and compared with explicit
results.

\section{Sum rule examples}
\label{sec:sum_rules}

The derivation of many energy weighted
sum rules has been succinctly described\cite{tinkham_2} as making use
of a `{\it...well-known technique which involves closure and evaluating
a double commutator in two different ways.}'  Such calculations rely
on the fact that the solutions of the system under study
form a complete set of states. For example, consider a system with energy eigenstates satisfying
$\hat{H} |n \rangle = E_n |n\rangle$. Then, for an arbitrary operator,
$\hat{O}$, we have the sum over off-diagonal matrix elements
\begin{eqnarray}
\sum_{all\, k} |\langle n | {\hat O}| k \rangle|^2
& = &
\sum_{all\, k} \langle n | \hat{O} | k \rangle
\langle k | \hat{O} | n \rangle
\nonumber\\
& = &
\langle n | \hat{O}
\left\{
\sum_{all\, k} |k\rangle \langle k |
\right\}
\hat{O} | n \rangle
\nonumber \\
& = &
\langle n | \hat{O}^2| n \rangle
\label{general_closure_sum_rule}
\,.
\end{eqnarray}
We note that the sum over the complete set of intermediate states, $|k\rangle$,
can include both an infinite sum (for discrete levels), an integral
(for continuum states), or both.

For the special case of $\hat{O} = x$, we obtain the simplest dipole
matrix element sum rule listed in Bethe and Jackiw,\cite{bethe_intermediate,jackiw_sum_rules} namely
\begin{equation}
\sum_{k} |\langle n |x | k \rangle|^2 = \langle n |x^2 | n \rangle
\qquad
\mbox{($x$-closure sum rule)}
\label{position_closure_sum_rule}
\end{equation}
with an identical sum rule relation for the off-diagonal matrix elements
of the momentum operator.

To derive the Thomas-Reiche-Kuhn (TRK) sum rule, we start with two
commutation relations,
\begin{equation}
[\,\hat{p},\,x\,] = \frac{\hbar}{i}
\quad
\mbox{and}
\quad
[\,\hat{H},\,x\,]
= \frac{1}{2m}[\,\hat{p}^2,\,x\,]
= \frac{\hbar}{mi} \, \hat{p}
\label{two_commutators}
\end{equation}
where we assume a standard 1D Hamiltonian of the form
\begin{equation}
\hat{H} = \frac{\hat{p}^2}{2m} + V(x)
\,.
\end{equation}
The first of these relations can be written in the form
\begin{equation}
\frac{\hbar}{i}
=
\langle n |\hat{p} x - x \hat{p}|n \rangle
=
\sum_{all\, k} \left\{
\langle n |\hat{p}|k\rangle
\langle k|x|n\rangle
-
\langle n |x |k\rangle
\langle k |\hat{p} |n\rangle
\right\}
\label{basic_one}
\end{equation}
where we have inserted a complete set of states. The second
relation in Eqn.~(\ref{two_commutators}) can be written as
\begin{equation}
\langle n |\hat{p}|k\rangle
= \frac{im}{\hbar} \langle n|[\,\hat{H},\,x\,]|k\rangle
= \frac{im(E_n-E_k)}{\hbar}
\langle n|x|k\rangle
\label{matrix_element_connection}
\end{equation}
with a similar expression for $\langle k |\hat{p}|n\rangle$. When
used in Eqn.~(\ref{basic_one}), this gives the desired result,
\begin{equation}
\frac{\hbar^2}{2m} = \sum_{k}(E_k-E_n)|\langle n|x|k\rangle|^2
\,.
\end{equation}

Wang\cite{wang_sum_rule} has derived a very general expression
for the energy-difference weighted sum rules for the matrix elements
of a well-behaved function of $x$, $F(x)$, namely
\begin{equation}
\sum_{k} (E_k - E_n) |\langle n|F(x)|k\rangle|^2
= \frac{\hbar^2}{2m}
\left\langle
n \left|
\frac{dF(x)}{dx}
\,
\frac{dF^{\dagger}(x)}{dx}
\right| n \right\rangle
\end{equation}
which simplifies if the function is Hermitian so that $F(x) = F^{\dagger}(x)$.
This general result can be used to immediately reproduce the TRK sum
rule by using $F(x) = x$. We can then also derive the Bethe sum
rule\cite{bethe_sum_rule} by using $\hat{O} = e^{iqx}$ in which
case we find
\begin{equation}
\sum_{k}(E_k-E_n)
|\langle n |e^{iqx}|k\rangle|^2 = \frac{\hbar^2 q^2}{2m}
\, .
\end{equation}
If we use $F(x) = x^2$, we obtain the so-called `monopole sum rule,'
which has been used in applications to nuclear collective excitations,\cite{bohigas}
\begin{equation}
\sum_{k} (E_k - E_n)|\langle n |x^2|k\rangle|^2 =
\frac{2\hbar^2}{m} \langle n |x^2|n \rangle
\, .
\label{monopole_sum_rule}
\end{equation}
Wang\cite{wang_sum_rule} also discussed sum rules involving
functions of the momentum operator, and `mixed' $x,\hat{p}$ relations.

Bethe and Jackiw\cite{bethe_intermediate,jackiw_sum_rules}
derive several other sum rules for dipole moment matrix elements
by using multiple commutation relations
with the Hamiltonian, thus generalizing Eqn.~(\ref{two_commutators}), and
yielding higher powers of the energy difference:

\begin{eqnarray}
\sum_{k} (E_k - E_n)^2 &&|\langle n |x|k \rangle|^2
=\frac{\hbar^2}{m^2} \langle n |\hat{p}^2|n\rangle\nonumber\\
&& =\frac{2\hbar^2}{m} \left\{E_n - \langle n|V(x)|n\rangle \right\}
\label{second_power_momentum_sum_rule}
\,.
\end{eqnarray}
\begin{equation}
\sum_{k}(E_k - E_n)^3 |\langle n |x| k \rangle|^2
= \frac{\hbar^4}{2m^2} \left\langle n \left|
\frac{d^2 V(x)}{dx^2} \right| n \right\rangle
\label{first_potential_sum_rule}
\end{equation}
and
\begin{equation}
\sum_{k} (E_k - E_n)^4 |\langle n |x| k \rangle|^2
= \frac{\hbar^4}{m^2} \left\langle n \left|\left(\frac{dV(x)}{dx}\right)^2
\right| n \right\rangle
\label{second_potential_sum_rule}
\end{equation}
where Eqns.~(\ref{first_potential_sum_rule}) and (\ref{second_potential_sum_rule}) are described as the
``{\it force times momentum}'' and ``{\it force squared}'' sum rules,
respectively.

We note that not all of these sum rules are guaranteed to converge\cite{bethe_intermediate}
and in our case, because of the singular nature
of the potentials used here as idealized models
(the infinite well and the single-$\delta$ cases)
several of these sum rules will not be applicable.

\section{The infinite square well}
\label{sec:isw}

The infinite square well (ISW) potential is the most frequently
presented of all textbook examples of bound state systems  and is
frequently used as a model system to introduce students to tractable
examples of research level physics, such as wave packet revivals.\cite{bluhm}
We can confirm many of the sum rules discussed above for this case, making
use of relatively straightforward mathematical techniques to evaluate
the infinite sums which appear. (The only example we can find in the
literature of the evaluation of sum rules in the context of the
infinite well is  a short discussion in Ref.~[14].)

We consider the standard ISW potential, defined by
\begin{equation}
V(x) =
\left\{
\begin{array}{cl}
0 & \mbox{for $0<x<a$} \\
\infty & \mbox{for $x<0$ and $x>a$}
\end{array}
\right.
\, .
\end{equation}
The energy eigenstates and corresponding eigenvalues are
\begin{equation}
\psi_n(x) = \sqrt{\frac{2}{a}}\, \sin\left(\frac{n\pi x}{a}\right)
\quad
\mbox{and}
\quad
E_n = \frac{\hbar^2 n^2 \pi^2}{2ma^2}
\end{equation}
where $n=1,2,...$ and the expectation value of $x^2$ required for the closure sum rule in
Eqn.~(\ref{position_closure_sum_rule}) is easily calculated to be
\begin{equation}
\langle n |x^2 |n \rangle  =  a^2\left(\frac{1}{3}
- \frac{1}{2 n^2\pi^2}\right)
\, .
\label{isw_xx}
\end{equation}
The energy differences needed for the various sum rule calculations
are given by
\begin{equation}
E_k - E_n = \frac{\hbar^2 \pi^2}{2ma^2} (k^2 - n^2)
\label{isw_energy_differences}
\,,
\end{equation}
while the off-diagonal matrix elements are given by
\begin{eqnarray}
\langle n |x|k \rangle
& = & \frac{2}{a}
\int_{0}^{a}
\,
\sin\left(\frac{n\pi x}{a}\right)
\, x \,
\sin\left(\frac{k\pi x}{a}\right)
\, dx
\nonumber \\
& = &
\left\{
\begin{array}{cc}
0 & \;\; \mbox{$k+n$ even} \\
-(8na/\pi^2)[k/(k^2-n^2)^2] & \;\; \mbox{$k+n$ odd}
\end{array}
\right .
\label{isw_matrix_elements}
\end{eqnarray}
so that for $n$ even (odd) only  odd (even) values of $k$ will contribute.
This result is due to the energy eigenfunctions' generalized parity property relative to
the center of the well at $x=a/2$. For the closure identity in Eqn.~(\ref{position_closure_sum_rule}),
we also need to include the diagonal matrix element,
\begin{equation}
\langle n |x| n \rangle = \frac{a}{2}\;.
\end{equation}
This term does not contribute to the other sum rules, since the
$k=n$ term is suppressed by the $(E_k-E_n$) energy difference factor.
In contrast to potential energy functions that are symmetric about the origin, such as the harmonic oscillator potential and the single $\delta$-function potential, the ISW potential as defined above is not symmetric and one must consider the $k=n$ case for the closure identity.

The position closure sum rule in Eqn.~(\ref{position_closure_sum_rule})
then reads
\begin{equation}
\sum_{all \, k} |\langle n |x|k \rangle|^2
= \left(\frac{a}{2}\right)^2
+
\left(\frac{8na}{\pi^2}\right)^2
\sum_{k} \frac{k^2}{(k^2-n^2)^4}
\label{first_isw_identity}
\end{equation}
where the summation is over even (odd) values $k$ if $n$ is odd (even).
This is the first of many examples we will encounter where we require
infinite summations of the form
\begin{equation}
S_p^{(\pm)}(z) = \sum_{k}\frac{1}{(k^2 - z^2)^p}
\end{equation}
where $z$ takes on integral values, and where the summation
is over odd, $S^{(-)}$, or even, $S^{(+)}$, values of $k$. For example,
the required summation in Eqn.~(\ref{first_isw_identity}) can be
written in the form
\begin{equation}
\sum_{k} \frac{k^2}{(k^2-n^2)^4}
=
\sum_{k} \frac{(k^2-n^2+n^2)}{(k^2-n^2)^4}
= S_{3}^{(\pm)}(n) + n^2 S_{4}^{(\pm)}(n)
\,.
\end{equation}
We provide a brief, but complete, review of how all of the sums
required in this section can be evaluated using standard series
expansions in Appendix~\ref{sec:summations}. We note, however,
that modern computer algebra systems (such as Mathematica)
can easily handle such sums.  Students may be allowed on first pass to use
such tools and then asked to delve more deeply into the methods used to
obtain the general mathematical results for this class of problems.

For example, in modified Mathematica syntax, the summation over
even integers $k$ (relevant for $n$ odd), yields:
\begin{verbatim}
Sum[k^2/(k^2-z^2)^4,{k,2,Infinity,2}]
= (-12 Pi Cot[Pi z/2] - 6 Pi^2 z Csc[Pi z/2]^2
+ 2 Pi^4 z^3 Cot[Pi z/2]^2 Csc[Pi z/2]^2
+ Pi^4 z^3 Csc[Pi z/2]^4)/768z^5
\end{verbatim}
so that for odd integer values of $z=n$, we have (by hand or by using \verb"Assuming-> z"$\;\in\;$\verb"Integers" in Mathematica)
\begin{equation}
\sum_{k\, even}
\frac{k^2}{(k^2-n^2)^4} =
\frac{\pi^4 n^3 - 6\pi^2 n}{768 n^5}
=
\frac{\pi^4}{768 n^2} - \frac{\pi^2}{128 n^4}
\, .
\label{obscure_sum}
\end{equation}
We obtain the same result (same function of $n$) for the
summation over odd values of $k$ (relevant for even $n$).  A
trivial modification (one character in fact) of the Mathematica
code is all that is required.  Using this result in Eqn.~(\ref{first_isw_identity}),
we then find that
\begin{eqnarray}
\sum_{all\, k} |\langle n |x|k\rangle|^2
& = &
|\langle n |x|n\rangle|^2
+
\sum_{k\neq n} |\langle n |x|k\rangle|^2
\nonumber \\
& = & \frac{a^2}{4}
+
\frac{64 a^2n^2}{\pi^4}
\left(\frac{\pi^4}{768 n^2} - \frac{\pi^2}{128 n^4}
\right)
\nonumber \\
& = & a^2 \left(\frac{1}{3} - \frac{1}{2n^2\pi^2}\right)
= \langle n |x^2 | n \rangle
\end{eqnarray}
as expected.

The TRK sum rule is then given by
\begin{eqnarray}
\sum_{k}(E_k-E_n) \, |\langle n|x|k\rangle|^2
& = &
\left(\frac{\hbar^2}{2m}\right)
\,
\left(\frac{64n^2}{\pi^2}\right)
\, \sum_{k}\frac{k^2}{(k^2-n^2)^3}
\label{trk_sum_rule_intermediate}
\end{eqnarray}
where the summation over $k$ is only for even (odd) values for
$n$ odd (even). These sums can also be done in closed form and one
finds
\begin{eqnarray}
{\cal I}_{n}^{(+)}(z)
& \equiv  &
\sum_{k\, \, even} \frac{k^2}{(k^2-z^2)^3}
= S_{2}^{(+)}(z) + z^2S_{3}^{(+)}(z)
\nonumber \\
& = &
\frac{1}{64z^3}
\left[
\pi^2 z \csc^2\left(\frac{\pi z}{2}\right)
+ 2\pi \cot\left(\frac{\pi z}{2}\right)
- \pi^3 z^2 \cot\left(\frac{\pi z}{2}\right) \cos^2\left(\frac{\pi z}{2}\right)
\right]
\label{even_identity} \\
{\cal I}_{n}^{(-)}(z)
& \equiv &
\sum_{k\, \, odd} \frac{k^2}{(k^2-z^2)^3}
= S_{2}^{(-)}(z) + z^2S_{3}^{(-)}(z)
\nonumber \\
& = &
\frac{1}{64z^3}
\left[
\pi^2 z \sec^2\left(\frac{\pi z}{2}\right)
- 2\pi \tan\left(\frac{\pi z}{2}\right)
+ \pi^3 z^2 \tan\left(\frac{\pi z}{2}\right) \sec^2\left(\frac{\pi z}{2}\right)
\right]
\label{odd_identity}
\end{eqnarray}
and we note the similarities in form. Inserting the appropriate
odd and even values of $n$, in each case we find that
\begin{equation}
{\cal I}_{n}^{(+)}(n) = {\cal I}_{n}^{(-)}(n)
= \frac{\pi^2}{64n^2}
\end{equation}
for all integral values of $n$. This result, when substituted into
Eqn.~(\ref{trk_sum_rule_intermediate}), directly confirms
the TRK sum rule.

Verification of the monopole sum rule in Eqn.~(\ref{monopole_sum_rule})
requires a small, but important modification of the summation methods.
The off-diagonal matrix elements required for
$k\neq n$ are
\begin{equation}
\langle n |x^2|k\rangle
= \frac{(-1)^{k-n}8a^2n}{\pi^2}\left(\frac{k}{(k^2-n^2)^2}\right)\, ,
\end{equation}
while for $k=n$,  one uses the result in Eqn.~(\ref{isw_xx}).
Since the $k=n$ term does not contribute to the sum
(because of the associated energy difference factor)
the left-hand side of Eqn.~(\ref{monopole_sum_rule}) reduces to
\begin{equation}
\sum_{k} (E_k-E_n)|\langle n|x^2|k\rangle|^2
= \left(\frac{\hbar^2\pi^2}{2ma^2}\right)
\left(\frac{64n^2a^4}{\pi^4}\right)
\sum_{k\neq n} \frac{k^2}{(k^2-n^2)^3}
\label{local_result}
\end{equation}
and we must sum over all values of $k\neq n$ since the even/odd pattern
seen in the dipole matrix elements is not present in this case.

In order to evaluate this sum, just as discussed in
Appendix~\ref{sec:summations},
we can first generalize the sum to non-integer values of $n$, and then
rewrite the sum as
\begin{equation}
T(z;n) \equiv
\sum_{k\neq n} \frac{k^2}{(k^2-z^2)^3}
=
\left[\,\sum_{all\, k} \frac{k^2}{(k^2-z^2)^3}\,\right]- \frac{n^2}{(n^2-z^2)^3}
\end{equation}
where the second term corresponds to the `missing' term in the $k\neq n$ summation.
The first sum can be evaluated for arbitrary $z$, giving the result
\begin{equation}
T(z;n) =
\left(\frac{\pi \cot(\pi z) + \pi^2 z\csc^2(\pi z)
- 2\pi^3 z^2 \cot(\pi z) \csc^2(\pi z)}{16z^3}\right)
- \frac{n^2}{(n^2-z^2)^3}
\,.
\end{equation}
Since we will be taking the limit where
$z \rightarrow n$ (and integral), we write $z = n +\epsilon$ for
general $n$, and we find that
both terms have factors which diverge as $1/\epsilon^3$, $1/\epsilon^2$, and $1/\epsilon$.
If, however, we expand both terms about $z=n$ (\textit{i.e.} in small values
of $\epsilon$) we find that these divergences cancel, leaving
the finite result
\begin{equation}
\lim_{z \rightarrow n} T(z;n)
= \lim_{\epsilon \rightarrow 0} T(n+\epsilon;n)
=
T(n) = \frac{\pi^2}{16n^2}\left( \frac{1}{3} - \frac{1}{2n^2\pi^2}\right)
\end{equation}
which when inserted into Eqn.~(\ref{local_result}) reproduces the right-hand side of
Eqn.~(\ref{monopole_sum_rule}).

Many of the other sum rules discussed in
Sec.~\ref{sec:sum_rules},
such as those that require derivatives of the potential energy function, Eqns.~(\ref{first_potential_sum_rule}) and (\ref{second_potential_sum_rule}),
are not well-defined for the infinite
square well (or the single $\delta$-function in Sec.~\ref{sec:delta})
due to the singular nature of the potential energy function.
While the matrix elements $\langle n|e^{iqx}|k\rangle$ required
for the Bethe sum rule in Eqn.~(\ref{bethe_sum_rule}) are easily
obtained in closed form, the summation methods discussed here are
not immediately applicable.

We can now use identical methods to evaluate the second-order shift
of the energy levels of the infinite square well due to the addition
of a linear potential, $V'(x) = Fx$, namely the Stark effect. In this
geometry, where the ISW potential is not symmetric, the first-order
energy shift is non-vanishing and is given by
\begin{equation}
E_n^{(1)} = \langle n |Fx|n \rangle
= \frac{aF}{2}
\,.
\end{equation}
The second-order shift has been evaluated by Mavromatis
for the ground state\cite{mavromatis_1} and then extended to a general
state\cite{mavromatis_2,mavromatis_3} by using variations
on the Dalgarno-Lewis method.\cite{dalgarno_and_lewis}
If we explicitly write the standard expression for the second-order
energy shift, we have
\begin{eqnarray}
E_n^{(2)} = \sum_{k \neq n}
\frac{|\langle n|Fx|k\rangle|^2}{(E_n^{(0)} - E_k^{(0})}
& = & - \left(\frac{F^2 2ma^2}{\hbar^2}\right)
\left(\frac{8na}{\pi^2}\right)^2
\sum_{k} \frac{k^2}{(k^2-n^2)^5}
\label{isw_stark_result}
\end{eqnarray}
which is formally identical to the class of summations discussed here.
Using either Mathematica or the results of Appendix~\ref{sec:summations}
we find that the required sum (for either $n$ even or odd) is given
by
\begin{equation}
\sum_{k} \frac{k^2}{(k^2-n^2)^5}=\frac{15\pi^2n - \pi^4 n^3}{3072 n^7}
\end{equation}
so that
\begin{equation}
E_n^{(2)} =
- F^2 \left(\frac{ma^4}{\hbar^2}\right)
\left(\frac{15 - (n\pi)^2}{24 \pi^2n^4}\right)
\,.
\end{equation}
The overall $n$-dependent form agrees with the results of
Mavromatis,\cite{mavromatis_1,mavromatis_2,mavromatis_3} who considered
the related problem of the symmetric infinite well, for which the
first-order correction vanishes. This result is interesting in itself
as the second order shift for the ground state
is negative (as it always should be, since all states contributing to
Eqn.~(\ref{isw_stark_result}) are higher in energy) but for $n=2$ and
higher, the shift changes sign. This is in contrast to the
behavior of the harmonic oscillator, where the second-order shift is
always negative, independent of quantum number.

\section{The single $\delta$-function potential}
\label{sec:delta}

Another popular model system in which to investigate sum rule and
perturbation theory results is the single (attractive) $\delta$-function
potential, defined here by
\begin{equation}
V_{\delta}(x) = - g\delta(x)
\,.
\label{delta_function_definition}
\end{equation}
The use of $\delta$-function potentials as simply soluble
models of potential barriers or wells has a long history in
quantum mechanics, going back at least to Kronig and Penney\cite{kronig_and_penney} who considered a
`{\it series of equidistant rectangular barriers}' and then
took the limit where the `{\it ...the breadth $b$ of these
barriers is made infinitely small and their height $V_0$
infinitely large...}' while not actually using the $\delta$-function
notation.

Morse and Feshbach\cite{morse_and_feshbach}
explicitly considered the form in Eqn.~(\ref{delta_function_definition}),
make note of the correct (dis)continuity condition on the energy eigenfunction
at the origin, namely
\begin{equation}
\psi'(0^{+}) - \psi'(0^{-}) = -\frac{2mg}{\hbar^2} \psi(0)
\, ,
\end{equation}
cite  it as being
`{\it useful in the study of nuclear forces},' and go on to
discuss the single bound state as well as scattering solutions.
Frost\cite{frost} considered single and multiple attractive $\delta$-function
potentials as models of `{\it hydrogen-like atoms},'
the hydrogen molecule-ion and more complex
systems. He was perhaps the first to explicitly comment
on the similarities of the energy eigenvalue
and eigenfunction for the single bound state of this system to the
ground state of the Coulomb problem.  Since then, single and multiple
$\delta$-function potentials have been widely used in model calculations
in both the pedagogical and research literature.\cite{lapidus_references}

We note that compared to the two other most widely used
simple 1D models,  the infinite well and harmonic oscillator, the
$\delta$-function potential has the advantage that it admits both
bound and continuum solutions, as does the Coulomb potential, and so
it presents new features compared to purely discrete spectra.

The single bound ($E<0$) state for the potential in
Eqn.~(\ref{delta_function_definition}) is given by
\begin{equation}
\psi_{0}(x) = \sqrt{K_0} \, e^{-K_0|x|}
\label{single_bound_state}
\end{equation}
where $K_0 = \frac{mg}{\hbar^2}$
with the corresponding bound state energy eigenvalue
\begin{equation}
E_0 = - \frac{mg^2}{2\hbar^2} = - \frac{\hbar^2K_0^2}{2m}
\,.
\label{single_delta_ground_state_energy}
\end{equation}
One can then note many comparisons to the ground state of the hydrogen atom,
if one defines the Coulomb potential as $V_c(r) = - \frac{g}{r}$ and one defines
and substitutes $a_0 \equiv \frac{1}{K_0}$ in Eqns.~(\ref{single_bound_state}) and (\ref{single_delta_ground_state_energy}).  Not only does the form of the ground state energy in
Eqn.~(\ref{single_delta_ground_state_energy}) match that of the Coulomb potential,
but the form of the energy eigenfunction in Eqn.~(\ref{single_bound_state}) does as well.

For use in confirming the closure relations in
Eqn.~(\ref{position_closure_sum_rule}),
we find that for
the ground state energy eigenfunction we have
\begin{equation}
\langle 0| x^2 | 0 \rangle
=  \frac{1}{2K_0^2}
\label{two_squared_deltas}
\,.
\end{equation}

The $E>0$ continuum states can be classified by their parity and are given by
\begin{eqnarray}
\psi_{k}^{(-)}(x) & = & \frac{1}{\sqrt{\pi}} \sin(kx)
\label{odd_continuum_states}
\\
\psi_{k}^{(+)}(x) & = & \frac{1}{\sqrt{\pi (k^2 + K_0^2)}}
\, \left(K_0 \sin(k|x|) - k \cos(kx)\right)
\label{even_continuum_states}
\end{eqnarray}
both of which have the same free-particle energy $E_k = \hbar^2k^2/2m$.
The combination of the single bound state in  Eqn.~(\ref{single_bound_state})
and the continuum states in Eqns.~(\ref{odd_continuum_states})
and (\ref{even_continuum_states}) have been explicitly shown\cite{delta_completeness}
to form a complete set of states. The effect
of the continuum states on a simple perturbation theory calculation
has also been demonstrated by Kiang.\cite{kiang}

For the various sum rules, we will consider here
the $|n\rangle = |0\rangle$ case only,
as others using purely continuum states do not converge.
Because of the symmetry of the system, parity arguments dictate that
the only non-zero dipole matrix elements connecting the single ground state
to the continuum will arise from the $\psi_{k}^{(-)}(x)$ states,
and we find that
\begin{equation}
\langle 0 |x| k^{(-)}\rangle
= \int_{-\infty}^{+\infty}
\left(\sqrt{K_0}e^{-K_0|x|} \right)
\, x \,
\left(\frac{1}{\sqrt{\pi}} \sin(kx)\right)
\, dx
= 4 \sqrt{\frac{K_0^3}{\pi}}
\, \frac{k}{(K_0^2 + k^2)^2}
\,.
\end{equation}
The energy differences are then given by
\begin{equation}
E_k - E_0 = \frac{\hbar^2}{2m} (k^2 + K_0^2)
\, .
\end{equation}
and we note the similarities in form between these two expressions
and the corresponding results for the ISW in
Eqns.~(\ref{isw_energy_differences})
and (\ref{isw_matrix_elements}).

The dipole matrix element closure relation
in Eqn.~(\ref{position_closure_sum_rule}) then
becomes
\begin{equation}
\sum_{k} |\langle 0 |x| k^{(-)}\rangle|^2
= \left(\frac{16K_0^3}{\pi}\right)
\,
\int_{0}^{\infty} \frac{k^2}{(k^2+K_0^2)^4}\, dk
= \frac{1}{2K_0^2}
= \langle 0 |x^2| 0 \rangle
\end{equation}
where the integral can be done by standard methods,
and agrees with the value
in Eqn.~(\ref{two_squared_deltas}).

The left hand-side of the TRK sum rule in Eqn.~(\ref{trk_sum_rule})
gives
\begin{equation}
\int_{0}^{\infty}  (E_k -E_0) |\langle 0 |x| k^{(-)} \rangle|^2\, dk
 =
\frac{\hbar^2}{2m} \left(\frac{16 K_0^3}{\pi}\right)
\int_{0}^{+\infty}
\frac{k^2}{(k^2 + K_0^2)^3}\,dk
= \frac{\hbar^2}{2m}
\end{equation}
also as expected. We note the similarity in form of these integral expressions to the summation results for the infinite well in Eqns.~(\ref{trk_sum_rule_intermediate})
and (\ref{first_isw_identity}).

In order to confirm the monopole sum rule in Eqn.~(\ref{monopole_sum_rule}),
we require the off-diagonal matrix elements of $x^2$ for which only the
even continuum states in Eqn.~(\ref{even_continuum_states})
contribute, giving
\begin{equation}
\left\langle 0 \left| x^2 \right| k^{(+)}\right\rangle
= 8 \sqrt{\frac{K_0^3}{\pi(k^2+K_0^2)}}
\, \frac{k}{(k^2+K_0^2)^2}
\,.
\end{equation}
We then find that
\begin{equation}
\sum_{k} (E_k-E_0)|\langle 0 |x^2| k^{(+)}\rangle|^2
= \left(\frac{\hbar^2}{2m}\right)
\left(\frac{64K_0^3}{\pi}\right)
\int_{0}^{\infty}
\frac{k^2}{(k^2+K_0^2)^4}\,dk
= \frac{\hbar ^2}{mK_0^2}
\end{equation}
and we note that factors of $(k^2+K_0^2)$ from the energy difference
and energy eigenfunction normalization in the numerator and denominator
respectively cancel.

The Stark effect for the single $\delta$-function potential has been
analyzed using exact results from the Airy function solutions,
\cite{russian_stark_delta,fernandez_stark_delta} as well as
using the Dalgarno-Lewis method.\cite{maize_stark_delta} Using the
dipole matrix elements derived above, we can evaluate the second-order
energy shift directly, using the same kinds of straightforward integrals
encountered so far. We find that
\begin{eqnarray}
E_0^{(2)}
& = & \int_{k}
\frac{|\langle 0 |Fx| k^{(-)}\rangle|^2}{(E_0^{(0)} - E_{k}^{(0)})}
\nonumber \\
& = &
- \frac{2mF^2}{\hbar^2}
\left(\frac{16K_0^3}{\pi}\right)
\,
\int_{0}^{\infty} \left\{\frac{1}{(k^2+K_0^2)}\right\}
\frac{k^2}{(k^2+K_0^2)^4} \,dk
\nonumber \\
\label{second_order_delta_stark}
\end{eqnarray}
which agrees with the results of Refs.~[40]-[42], when put into this notation.
We note that the entire contribution to the Stark shift in the
ground state energy in this
case comes from the continuum states and this result is one of the
few examples of the explicit evaluation
of the contribution of the continuum terms in such a calculation.

We recall that for the hydrogen atom ground state,
the total second-order shift\cite{more_textbook} can be written in the form
\begin{equation}
E_0^{(2)}(\mbox{H-atom})
= - \frac{9}{4} \left(\frac{F^2a_0^3}{g}\right)
\label{total_hatom_stark}
\end{equation}
using the form of the Coulomb potential and perturbation theory,
this result comes from summing over the contributions of both
the bound states and continuum states. Ruffa\cite{ruffa}
has evaluated the continuum contribution to the expression in
Eqn.~(\ref{total_hatom_stark}) in terms of a single integral
and finds a net contribution of $0.4184$ to the total $9/4 = 2.25$
value of the pre-factor.
It is perhaps then fairer to compare the second-order Stark result
in Eqn.~(\ref{second_order_delta_stark}), namely $5/8 = 0.625$, to
that partial contribution.

Finally, the Bethe sum rule is given by
\begin{equation}
{\cal B} = \sum_{k} (E_k - E_0) \left|\langle 0  | e^{iqx} | k \rangle\right|^2
= \frac{\hbar^2 q^2}{2m}
\end{equation}
and in this case we will have
two contributions to the left-hand-side, coming from the
even ($e$) or odd ($o$) continuum states, namely
\begin{eqnarray}
{\cal B}
& = &
{\cal B}_{e} + {\cal B}_{o}
\nonumber \\
& \equiv &
\int_{0}^{\infty} (E_k-E_0)\left|\langle 0 | \cos(qx)| k^{(+)}\rangle\right|^2 \, dk
\nonumber \\
& & \qquad\qquad
+
\int_{0}^{\infty} (E_k-E_0)\left|\langle 0 | \sin(qx)| k^{(-)}\rangle\right|^2 \, dk
\end{eqnarray}
and we consider each term separately.
The first matrix element of interest is
\begin{eqnarray}
{\cal I}_{o} & = & \langle 0|\sin(qx)| k^{(-)}\rangle
\nonumber \\
& = & \sqrt{\frac{4K_0}{\pi}}
\int_{0}^{\infty} e^{-K_0x} \, \sin(qx)\, \sin(kx)\,dx
\nonumber \\
& = &
\sqrt{\frac{4K_0}{\pi}}
\left[
\frac{2kqK_0}{[(k+q)^2+K_0^2][(k-q)^2 + K_0^2]}
\right]
\end{eqnarray}
where we use the symmetry of the energy eigenfunctions to evaluate the
integral over positive values of $x$ only.
Recalling that $E_k-E_0 = \hbar^2(k^2+K_0^2)/2m$, we find
\begin{equation}
B_{o}
=
\frac{\hbar^2 q^2}{2m} \left(\frac{16K_0^3}{\pi}\right)
\,
\int_{0}^{\infty}
\,
\frac{k^2(k^2+K_0^2)}{[(k+q)^2 + K_0^2]^2[(k-q)^2 + K_0^2]^2}
\, dk
\,.
\label{single_delta_bethe_1}
\end{equation}
Use of an integrated mathematics package (again Mathematica) returns
the correct value for the integral, provided one correctly interprets
the many cautionary restrictions on the values of $K_0$ and $q$.  Given the
relatively complicated nature of the intermediate results coming from
such programs, however, it is again important to be able to check the expressions `by hand.' In this
case, it simply involves extending the integral over the entire real line
(since the integrand is an even function of $k$)
and then using contour integration methods
(see Appendix~\ref{sec:contour_integrals} for details), giving
\begin{equation}
{\cal B}_{o} = \left(\frac{\hbar^2 q^2}{2m}\right)
\left\{ \frac{K_0^2 + q^2/2}{K_0^2 + q^2}\right\}
\,.
\label{odd_bethe_contribution}
\end{equation}
For the even case, we require the matrix element
\begin{eqnarray}
{\cal I}_{e} & = &
\langle 0 |\cos(qx)| k^{(+)}\rangle
\nonumber \\
& = &
\sqrt{\frac{4K_0}{\pi (K_0^2+k^2)}}
\,
\int_{0}^{\infty}  \, e^{-K_0x}\,\cos(qx)\,
\left[K_0 \sin(kx) - k \cos(kx)\right]
\, dx
\nonumber \\
& = &
\sqrt{\frac{4K_0}{\pi (K_0^2+k^2)}}
\left[\frac{- 2kK_0 q^2}{[(k+q)^2+K_0^2][(k-q)^2 + K_0^2]} \right]
\end{eqnarray}
and the even contribution to the sum rule becomes
\begin{eqnarray}
{\cal B}_{e} & = &
\int_{0}^{\infty}
\, (E_k-E_0) |\langle 0 |\cos(qx)| k^{(+)}\rangle |^2
\, dk
\nonumber \\
& = &
\left(\frac{\hbar^2q^2}{2m}\right)
\left(\frac{ 8K_0^3 q^2}{\pi}\right)
\int_{-\infty}^{+\infty} \frac{k^2}{[(k+q)^2+K_0^2]^2[(k-q)^2 + K_0^2]^2}
\, dk
\,.
\label{single_delta_bethe_2}
\end{eqnarray}
The integral can again be done with similar contour methods giving the
result
\begin{equation}
{\cal B}_{e} = \left(\frac{\hbar^2 q^2}{2m}\right)
\left\{\frac{q^2/2}{K_0^2 + q^2}\right\}
\end{equation}
which can be combined with Eqn.~(\ref{odd_bethe_contribution}) to give
\begin{equation}
{\cal B} = {\cal B}_{o} + {\cal B}_{e} = \frac{\hbar^2 q^2}{2m}
\end{equation}
as expected.

\section{Conclusions and discussion}

We have presented an array of familiar (and not-so-familiar) one-dimensional
sum rules, a number of which have proved useful in the
development of many fields of physics. Using two standard
model systems as
testbeds, we have illustrated the diverse ways in which such sum rules are
confirmed, emphasizing the different mathematical techniques
(infinite summation tricks and contour integration methods) used
in each case. While the evaluation of the necessary summations or integrals
can be simplified by the use of integrated mathematics programs, we
have also provided the details necessary to demonstrate the same results
from first principles.

We have also noted the striking similarities
of some of the expressions which arise for the same sum rules in the
infinite square well and single $\delta$-potential cases.
Despite the qualitatively very different physical behavior of the two
systems, they both begin with free-particle solutions. The infinite wall
boundaries of the ISW force quantized eigenstates with $E_n
= \hbar^2 k_n^2/2m$, while the attractive $\delta$-function gives the
identical dispersion relation for the $E>0$ states, but
with continuous $k$-values.  The $\delta$-function case also includes one
$E<0$ state for which the sinusoidal solution is analytically continued
to the localized exponential form in Eqn.~(\ref{single_bound_state}).
The connections between these two model systems are seldom if ever
stressed, but appear very naturally in these sum rule calculations.

We hope that the suite of exemplary problems discussed here
can be useful to instructors in lectures as well as for
homework problems, in both the advanced undergraduate and graduate
quantum mechanics curriculum, especially by putting this important
tool of theoretical physics into a historical and research context.

\appendix

\section{Infinite sums for the square well problem}
\label{sec:summations}

Many of the sum rule and second-order perturbation
theory results in Sec.~\ref{sec:isw} for the infinite square well
involve the evaluation of infinite sums of the forms
\begin{equation}
S_{p}^{(+)}(z) = \sum_{even\, k}\frac{1}{(k^2-z^2)^{p}}
\qquad
\quad
\mbox{or}
\quad
\qquad
S_{p}^{(-)}(z) = \sum_{odd\, k}\frac{1}{(k^2-z^2)^{p}}
\label{required_summations}
\end{equation}
where the both expressions are eventually evaluated using
integral values of $z=n$,  with $n$ odd and even respectively so no divergences
occur. While multi-purpose
computer programs such as Mathematica can recognize and correctly evaluate
such sums, it can be important for some students (and many instructors)
to also be able to derive them `{\it from scratch}.'
To that end, in this Appendix
we provide a very brief, but self-contained and complete, review of
the mathematical tools necessary for their derivation from more basic
results with which students at this level should be quite familiar.

We begin by considering the general expression
\begin{equation}
S_{p}(z) \equiv \sum_{k=1}^{\infty} \frac{1}{(k^2-z^2)^{p}}
\end{equation}
where the summation is over all positive integer values of $k$.
The basic result we require is for the $p=1$ case,  namely
\begin{equation}
S_1(z) = \sum_{k=1}^{\infty}
\frac{1}{(k^2-z^2)}
=
\frac{1}{2z^2} - \frac{\pi \cot(\pi z)}{2z}
\label{handbook_result}
\end{equation}
which appears, for example, in Gradshteyn and Ryzhik. \cite{gr}
This standard `handbook' result can, in turn, be derived at a more
fundamental level from a Fourier series expansion\cite{mathews_and_walker}
by evaluating the Fourier components of the expansion
\begin{equation}
\cos(zx) = \frac{a_0}{2} + \sum_{n=1}^{\infty}
\left[a_n \cos(n x) + b_n \sin(n x)\right]
\end{equation}
over the interval $(-\pi, +\pi)$; note that here $z$ is considered a constant.
The Fourier coefficients can be
evaluated using standard integrals and we obtain
\begin{equation}
\cos(zx)
= \frac{\sin(z\pi)}{z\pi}
- \sum_{n=1}^{\infty}
\left[\frac{2z\sin(\pi z)\cos(n\pi)}{\pi (n^2-z^2)}\right]\, \cos(nx)
\end{equation}
since the $b_n=0$ by symmetry. If we then specialize to
$x = \pi$, and use the fact that $\cos^2(n\pi) = 1$, we find
\begin{equation}
\pi \cot(\pi z) = \frac{1}{z}
- 2z \sum_{n=1}^{\infty} \frac{1}{(n^2-z^2)}
\end{equation}
and we note that this partial fraction expansion of $\cot(\pi z)$
correctly encodes the information on the divergences of the function
at all integral (positive, negative, and zero) values of $z$.
Rewriting this expression, we find that
\begin{equation}
S_{1}(z) \equiv \sum_{n=1}^{\infty} \frac{1}{(n^2-z^2)}
= \frac{1}{2z}\left(\frac{1}{z} - \pi \cot(\pi z)\right)
= \frac{1}{2z^2} - \frac{\pi \cot(\pi z)}{2z}
\label{confirm_handbook}
\end{equation}
confirming the handbook result of Eqn.~(\ref{handbook_result}).
Such sums are already useful in that they can be used to
evaluate quantities such as the Riemann zeta function, defined by
\begin{equation}
\zeta(s) \equiv \sum_{n=1}^{\infty}\frac{1}{n^{s}}
\qquad
\quad
\mbox{giving}
\quad
\qquad
\zeta(2) = S_1(z=0) = \sum_{n=1}^{\infty}\frac{1}{n^2} = \frac{\pi^2}{6}
\end{equation}
as the $z\rightarrow 0$ limit of Eqn.~(\ref{confirm_handbook}).
(Such results can be directly connected to integrals which appear
frequently in the evaluation of quantities related to blackbody radiation
which students encounter in standard textbooks\cite{reif} on
statistical mechanics and can therefore be reinforced through such examples.)

If we differentiate the result in Eqn.~(\ref{confirm_handbook})
with respect to $z$, we find that
\begin{equation}
\frac{d}{dz} S_1(z) = 2z\, \sum_{k=1}^{\infty} \frac{1}{(k^2-z^2)^2}
= 2zS_2(z)
\end{equation}
so that in general
\begin{equation}
S_{p+1}(z)
= \frac{1}{2z} \frac{dS_{p}(z)}{dz}
\label{differentiation_trick}
\end{equation}
thereby generating sums of arbitrarily high power.  For example,
this gives
\begin{equation}
S_2(z) =
\frac{
\csc^2(\pi z)
[-2 + 2\pi^2 z^2 + 2\cos(2\pi z) + \pi z\sin(2\pi z)]}{8z^4}
\end{equation}
implying that $\zeta(4) = S_{2}(z=0) = \pi^4/90$.

Since our interest is often in summations restricted to the even or odd
integers, we write
\begin{equation}
S_1(z) =
\sum_{k=1}^{\infty} \frac{1}{(k^2-z^2)}
=
\sum_{k\,\,even}^{\infty} \frac{1}{(k^2-z^2)}
+
\sum_{k\,\,odd}^{\infty} \frac{1}{(k^2-z^2)}
\equiv
S_{1}^{(+)}(z) + S_{1}^{(-)}(z)
\, .
\end{equation}
We then note that
\begin{eqnarray}
S_{1}^{(+)}(z)
& = &
\sum_{k\,\,even}^{\infty} \frac{1}{(k^2-z^2)}
 = \sum_{l=1}^{\infty}
\frac{1}{((2l)^2 - z^2)}
\nonumber \\
& = & \frac{1}{4} \sum_{l=1}^{\infty}
\frac{1}{(l^2 - (z/2)^2)}
=  \frac{1}{4} S_{1}\left(\frac{z}{2}\right)
\nonumber \\
& = & \frac{1}{2z^2} - \frac{\pi \cot(\pi z/2)}{4z}
\label{even_sum}
\end{eqnarray}
which then gives
\begin{equation}
S_{1}^{(-)}(z) = S_{1}(z) - S_{1}^{(+)}(z)
= \frac{\pi}{4z}\left[\cot\left(\frac{\pi z}{2}\right) - 2\cot(\pi z) \right]
= \frac{\pi \tan(\pi z/2)}{4z}
\label{odd_sum}
\end{equation}
where we use half-angle formulae in the last step. Both of the
expressions in Eqns.~(\ref{even_sum}) and
(\ref{odd_sum}) can, of course, be confirmed using Mathematica.

The sums over higher powers of even/odd values of $n$ required
to evaluate $S_{p}^{(+)}(x)$ and $S_{p}^{(-)}(x)$
in Eqn.~(\ref{required_summations}) are then obtained by repeated
use of the differentiation trick in Eqn.~(\ref{differentiation_trick}).
For example, we obtain results such as
\begin{equation}
S_{2}^{(+)}(z)  =
\frac{\csc^2(\pi z/2)[-4 + \pi^2 z^2 + 4\cos(\pi z) + \pi z \sin(\pi z)]}{16x^4}
\end{equation}
and
\begin{equation}
S_{2}^{(-)}(z)  =
\frac{\pi \sec^2(\pi z/2)[\pi z - \sin(\pi z)]}{16 z^3}
\,.
\end{equation}

\section{Contour integrals}
\label{sec:contour_integrals}

The explicit evaluation of the integrals in Eqns.~(\ref{single_delta_bethe_1})
and (\ref{single_delta_bethe_2}) by contour integration techniques
can be done by extending the region
of integration over the entire real-axis. A contour consisting of a
semi-circle of radius $R$ can be then used as the
integrands both have simple (double) poles at $z_{0}^{(\pm)} = \pm q
+ iK_0$ in the upper-half plane.
The contribution to the contour integral over the circular arc
vanishes as $R\rightarrow \infty$, leaving
\begin{equation}
\int_{-\infty}^{+\infty} F(k)\,dk = 2\pi i \sum_{i} {\cal R}_i
\end{equation}
where the residues are given by
\begin{equation}
{\cal R}_i = \frac{1}{(n-1)!}
\left\{
\left(\frac{d}{dz}\right)^{n-1} [(z-z_0^{(i)})^n F(z)]
\right\}_{z \rightarrow z_0^{(i)}}
\end{equation}
for $z_0^{(i)} = z_0^{(\pm)}$ and where in this case $n=2$.

\end{document}